# RENEWABLE ENERGY LABORATORY FOR LIGHTING SYSTEMS

DUMITRU Cristian, "Petru Maior" University of Tg.Mureş
GLIGOR Adrian, "Petru Maior" University of Tg.Mureş

**ABSTRACT**
Nowadays, the electric lighting is an important part of our lives and also represents a significant part of the electric power consumption. Alternative solutions such as renewable energy applied in this domain are thus welcomed. This paper presents a workstation conceived for the study of photovoltaic solar energy for lighting systems by students of power engineering and civil engineering faculty.
The proposed system is realized to study the generated photovoltaic solar energy parameters for lighting systems. For an easier way to study the most relevant parameters virtual instrumentation is implemented. National Instruments' LabWindows / CVI environment is used as a platform for virtual instrumentation. For future developments remote communication feature intends to be added on which currently remote monitoring of solar photovoltaic energy and electric energy parameters are monitored.

## 1. INTRODUCTION

Solar energy is one of the renewable energies on which nowadays scientists interested in energy sources are focused. Considering the significant level of lighting energy consumption: home 25%, commercial buildings, schools, businesses 60% [1], renewable energy represents a viable alternative applicable in saving money spent on energy bills but also in reducing green house effect caused by emissions of classical power plants.

The efficient use of system based on solar energy involves the knowledge of solar datum and equipment characteristics from a studied area. By didactical point of view it is very handful allowing students to access real datum and concentrate on phenomenology using adequate tools. Authors of this paper have developed such a tool presented below.

## 2. SOLAR PHOTOVOLTAIC ENERGY DATUM AND CONVERSION

As the photovoltaic (PV) technology is an environmentally friendly method of generating electric power energy it represents an alternative to the present more and more expensive energy sources and therefore its study is an important key for future specialists. Following we will mention some characteristics that must be taken into consideration by those interested in study and implementation of photovoltaic solar energy, so that represents a preliminary approach before starting the so-called study on proposed workstation.

The solar energy availability depends on season, weather conditions, pollution or geographical position as Figure 1a illustrates for Tîrgu Mureş City [13].

Once a system is set up, the user only needs to monitor the proper operation and to check the energy yield. Regular checks ensure that the system delivers optimum yields and thereby achieves high efficiency levels. The user should check and record the meter readings and any special events for early recognition of faults and their cause [3].

Autonomous renewable systems, such as solar energy powered networks, experience real-time variations of input energy. By taking into account the variations of the load, control methods are required to maintain stability and achieve maximum penetration from the solar resource.

Considering Tîrgu Mureş as location, in Figure 1b is illustrated the (estimated) amount of electric power that can be expected each month from a PV system studied in this paper [13].

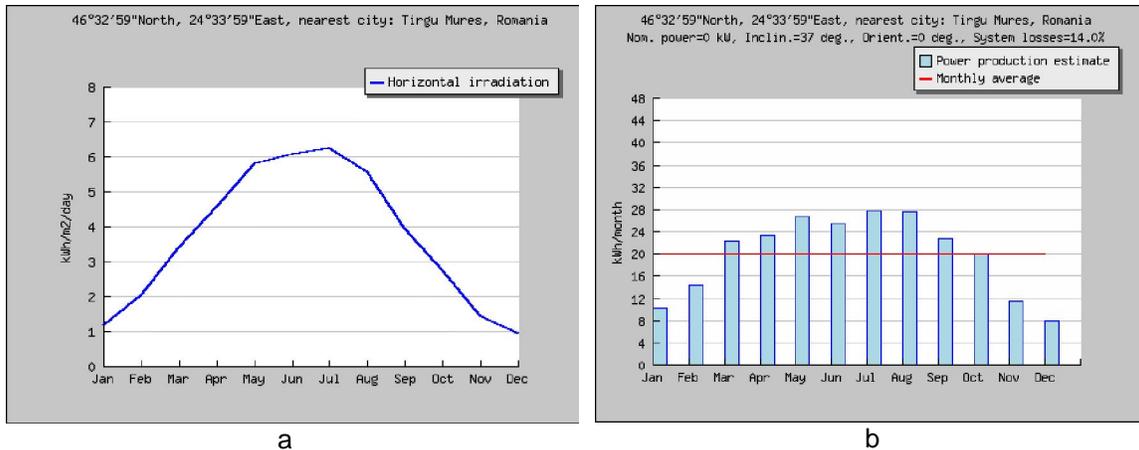

**Figure 1** (a) Monthly solar energy availability in Tîrgu Mures; (b The monthly amount, the expected average monthly and yearly production of electric power in Tîrgu Mures for a 0,2 kW solar panel [13]

### 3. PHOTOVOLTAIC LABORATORY
The proposed workstation for the study of lighting system based on photovoltaic solar energy is represented by a basic Lucas-Nulle workstation extended with monitoring system based on virtual instrumentation and possibility to study different lighting sources.
The workstation contains two types of solar energy converters: solar cells and photovoltaic solar panel.
Solar cells are represented by integrated SO 32107 M Lucas-Nulle modules. There are two types of solar cells: 2 modules of 6 Volts which can be connected serial or parallel configurations, and a 3 Volts module.
Solar panel is of ET-M53620 type characterized by the following parameters:
   - Peak power ($P_{max}$) 20 W
   - Maximum power point voltage ($V_{mpp}$) 17.82 V
   - Maximum power point current ($I_{mpp}$) 1.15 A
   - Open circuit voltage ($V_{oc}$) 21.96 V
   - Short circuit current ($I_{sc}$) 1.27 A
Energy conversion at a level usable by consumers is realized with two 12VDC-230VAC-50 Hz-150W inverters supplied by two solar charge controller SOLSUM 5.6 and SOLAR 18122. Supplemental energy is stored in a 12 V accumulator.
For lighting sources the workstation is adapted to host DC and AC lamp types.
On DC side primarily it can be found an incandescent lamp, a halogen lamp, a LED at 12V and a 12 V socket for connecting other type of lamps. The inverter offers the possibility to connect different types of AC lighting sources. For testing purposes is used as standard equipment a 230V/150W LEDs lamp.
For time analysis of solar energy availability a MacSolar device from Solarc is used.
The parameters of produced electric energy are monitored and then studied with virtual instrumentation module built by authors. This module consists from two parts: a hardware part for data acquisition and a software part for data analysis.

### 4. PARAMETERS MONITORING USING VIRTUAL INSTRUMENTATION
The presence of multiple parameters to be measured and recorded from one or more renewable energy sources or from one or more consumers leads to necessity of using computing systems. Therefore nowadays trend is to use on a large scale the virtual instrumentation.
Virtual Instrumentation represents the use of customizable software and modular measurement hardware to create user-defined measurement systems, consisting by virtual instruments.
Unlike the traditional measurement instruments, the virtual instruments do not have hard-coded functions; these systems are less limited in their versatility than virtual instrumentation systems. The primary difference between classical instrumentation and virtual instrumentation is the software component of a virtual instrument. The software enables complex and expensive equipment to be replaced by simpler and less expensive hardware. [4]

## 5. ARCHITECTURE OF THE SYSTEM
The main objective of proposed system is to provide the status of the solar panel and consumers parameters.
The architecture of the proposed system is presented in Figure. 2.

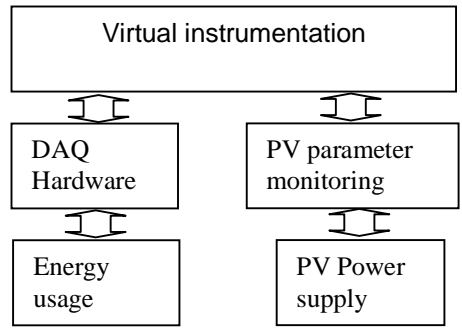

**Figure 2.** Architecture of the laboratory workstation

In the proposed architecture, the virtual instrumentation allows to preset the used devices and to view values of significant parameters. All data can be stored for future references and analysis.
The DAQ hardware purpose is the data acquisition. An ADVANTECH PCL711 DAQ card connected to a PC represents this module.
PV parameter monitoring module on hardware side consists from a MacSolar device.
Energy usage involves all lighting sources and PV power supply block refers to the main source of energy, PV ET-M53620 panel type.

## 6. HARDWARE IMPLEMENTATION OF THE MONITORING MODULE
In Figure 3 is presented the data acquisition system for monitoring of the solar energy parameters.
The acquisition system is built around a computer that obtains through a serial bus from the MACSOLAR electronic device the parameters available at solar cells and solar panel level.
In Figure 4 is presented the data acquisition system for monitoring of the electric solar energy parameter from solar panel output circuits.
This part is composed by a data acquisition board (PCL711) placed in a IBM compatible computer and a device (Figure 5) which adapts the voltage and current signals to the input of DAQ board.

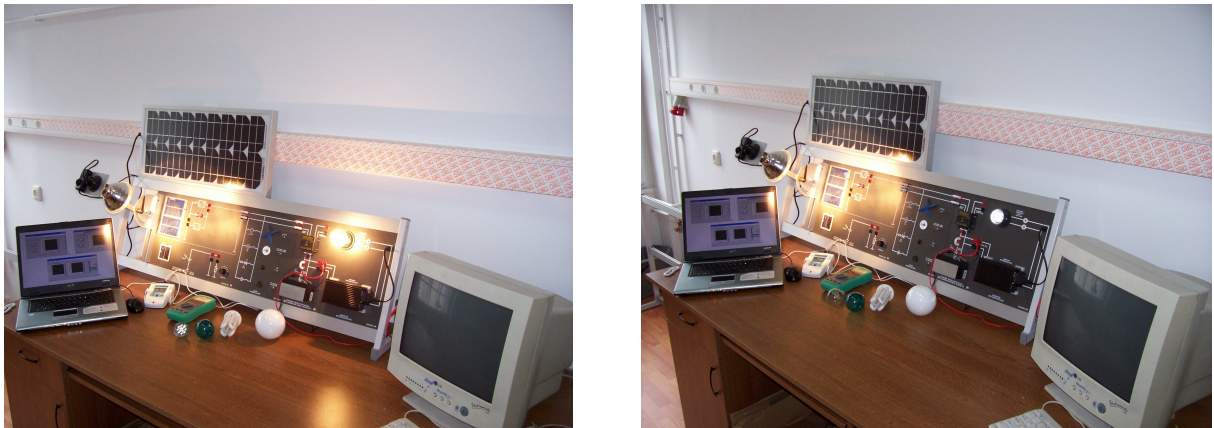

**Figure 3.** Experimental workstation for study of lighting system based on renewable energy

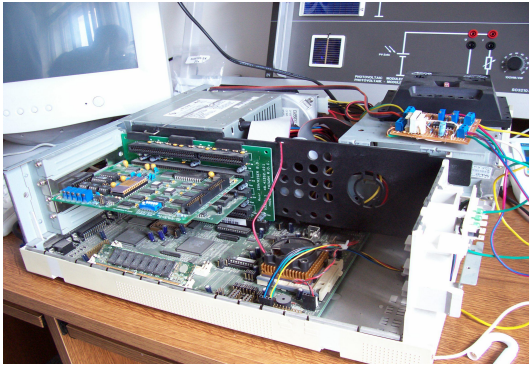

**Figure 4.** Hardware part of the acquisition system configured for monitoring of the electric solar energy parameter from solar panel output circuits

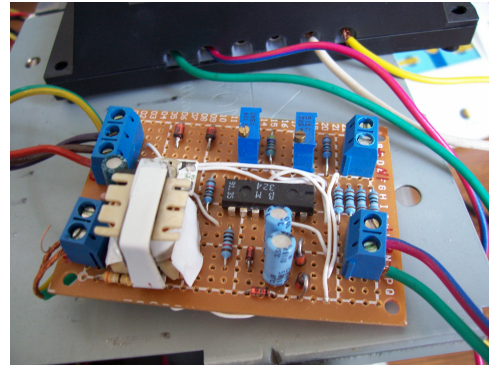

**Figure 5.** Conditioning device of the DAQ system configured for monitoring of the electric solar energy parameter from solar panel output circuits

## 7. SOFTWARE DEVELOPMENT FOR VIRTUAL INSTRUMENTATION

Software development was realized using National Instruments' LabWindows / CVI. It can be noted that this programming environment has been chosen because of its facility to develop applications in a productively manner allowing also to focus mostly on implementing the computation methods. By didactical point of view it is very handful allowing students to concentrate more on phenomenology with basics programming knowledge.

Software architecture (Figure 6) is based on 4 modules: acquisition module (AM), processing module (PM), communication module (CM) and user interface module (UI).

The AM module implements the function for setting-up the DAQ board and acquisition control.

PM module is designed for data processing and conditioning.

CM module assures the support for network communication.

In the UI modules are grouped the visual command panels with virtual instruments.

It must be noted that currently CM module is implemented with the support for DataSocket that is a data network programming technology, which simplifies data exchange between computers and applications. By using this technology, data from photovoltaic systems is accessed and shared from remote locations allowing the use of the workstation in a remote e-learning system or remote laboratory. In industrial application when distributed photovoltaic systems are available, data can be accessed and shared from remote locations for the final goal of proper system operation [5].

In Figure 7 is presented a generic sequence diagram of the data acquisition software module operation, where it can be observed the main object of the software modules and theirs interactions.

Figure 8 shows the main window of developed application. The window contains virtual instruments for voltage, current and power measurements.

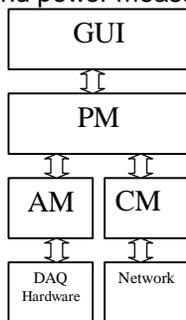

**Figure 6.** Main architecture of the software

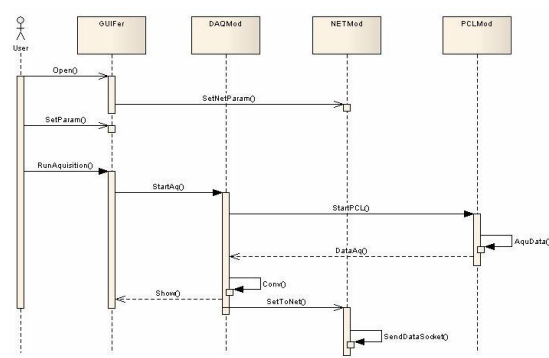

**Figure. 7.** Generic sequence diagram of the data acquisition software module operation

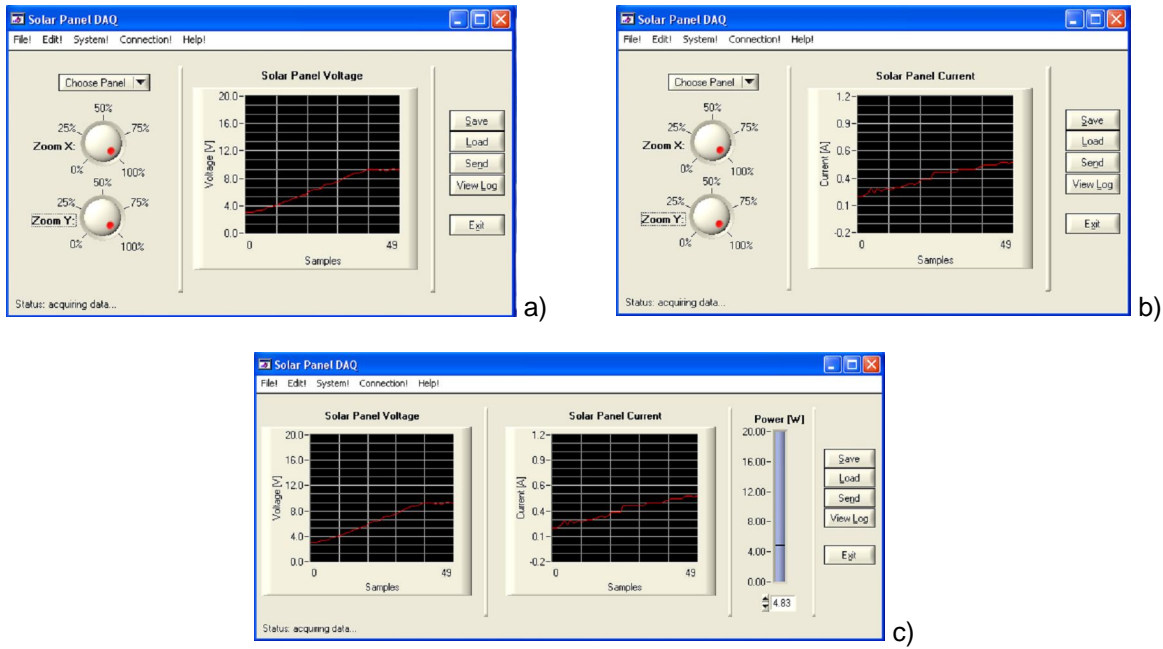

**Figure 8.** Solar panel a) current, b) voltage and c) power measurements window

### 8. CONCLUSIONS AND FUTURE WORK
In this paper was presented a laboratory workstation based on virtual instrumentation for study of lighting system supplied by photovoltaic solar energy. A few types of measurements and studies can be performed: measurement of irradiance, open circuit voltage and short-circuit current from photovoltaic cells, study of power, temperature, voltage and current characteristics from the power cells or modules, study of serial and parallel connection of solar modules and the partial shading of a photovoltaic generator, study of lighting sources directly fed or fed by stored power.
As a future work we intend to develop this application as a virtual laboratory accessible over computer network to study lighting system supplied by renewable energy.
Experimental stand presented can be used by students in: energy using laboratory and renewable energy laboratory for study of artificial lighting systems supplied by renewable energy.

### REFERENCES

1. ***, Intermediate Energy Infobook, Energy Consumption, 2008, http://www.need.org/, p. 44
2. Tom Markvart, Luis Castaner – Practical Handbook of Photovoltaics, Fundamentals and Applications, Ed. Elsevier, Oxford, UK, 2003
3. Sorensen, B. (2004). Renewable Energy: Its physics, engineering, use, environmental impacts, economy and planning aspects. Third Edition, Elsevier Academic Press, London.
4. *** http://en.wikipedia.org/wiki/Virtual_instrumentation
5. Snyman, D.B., Enslin, J.H.R. – Simplified maximum power point controller for PV installations, Photovoltaic Specialists Conference, 1993, Conference Record of the Twenty Third IEEE Volume, Issue, 10-14 May 1993 Page(s): 1240 – 1245
6. Basheer Ahmed, S., Cohen, M.J., Schweizer, T.C. – Renewable Energy Sources: Technology and Economics, IEEE, 1994
7. *** - Photovoltaic Solar Panel from Viessmann – The Right Solution for Every Application, For Generating Power from the Sun, Technical Guide, Vitovolt , 5/2004
8. *** - Photovoltaic Solar Panel from Viessmann – The Right Solution for Every Application, For Generating Power from the Sun, Datasheet, Vitovolt , 5/2004

9.  Dumitru, C.D. - Management of a System Based on Renewable Resources for Electric Power Supply of an Administrative Building. 1st International Conference on Modern Power Systems MPS 2006, pag.101-106
10. Whitaker, C.M. ş.a. - Application and Validation of a New PV Performance Characterization Method, The 26h IEEE Photovoltaic Specialists Conference, Anaheim, California, 1997.
11. \*\*\* Lab Windows/CVI User Manual, National Instruments, 2003 Edition;
12. \*\*\* PCL-711 B, PC-MultiLab CARD, USER'S MANUAL, Advantech Co., Ltd.
13. \*\*\* <http://re.jrc.ec.europa.eu/>
14. G.R. Walker (2000) "Evaluating MPPT converter topologies using a MATLAB PV model," Proceedings of the Australasian Universities Power Engineering Conference, AUPEC'00, Brisbane
15. \*\*\*, www.ni.com

**ACKNOWLEDGEMENTS**

Authors:
Ph.D student Cristian DUMITRU, assistant
"Petru Maior" University from Tg. Mureş
1, N. Iorga St., RO-540088 Targu-Mures, Romania
Ph.:+40.265.233112
Fax:+40.265.233212
e-mail: cdumitru@engineering.upm.ro

dr. Adrian GLIGOR, lecturer
"Petru Maior" University from Tg. Mureş
1, N. Iorga St., RO-540088 Targu-Mures, Romania
Ph.:+40.265.233112
Fax:+40.265.233212
e-mail: agligor@upm.ro